        \documentstyle[11pt,amsmath,amsfonts,euscript]{article}
        \newcommand{\al}{\alpha}
        \newcommand{\gam}{\gamma}
        \newcommand{\del}{\delta}
        
        \newcommand{\kap}{\kappa}
        \newcommand{\lam}{\lambda}
        \newcommand{\sig}{\sigma}
        \newcommand{\th}{\theta}
        \newcommand{\vth}{\vartheta}
        
        \newcommand{\Del}{{\mathit{\Delta}}}
        \newcommand{\Gam}{{\mathit{\Gamma}}}
        \newcommand{\Lam}{{\mathit{\Lambda}}}

        \newcommand{\HH}{{\cal H}}

        \newcommand{\SS}{{\cal S}}
        \newcommand{\TT}{{\cal T}}
        \newcommand{\GG}{{\EuScript{G}}}
        \newcommand{\MM}{{\EuScript{M}}}
        \newcommand{\ZZ}{{\EuScript{Z}}}

        \newcommand{\g}{{\mathfrak g}}
        \newcommand{\gt}{{\mathfrak t}}
        
        \newcommand{\ZA}{{\mathbb{Z}}}
        \newcommand{\PP}{{\mathbb{P}}}
        \newcommand{\QQ}{{\mathbb{Q}}}
        \newcommand{\RE}{{\mathbb{R}}}
        \newcommand{\CO}{{\mathbb{C}}}

        \newcounter{sect}
        \newcounter{subsect}
        \newcommand{\sect}[1]{\vspace{2ex}
                \addtocounter{sect}{1}\setcounter{subsect}{0}
                \begin{flushleft}
                {{\large\bf \arabic{sect}. {#1}}}
                \end{flushleft}
                \setcounter{thm}{0}
                \setcounter{equation}{0}
                \def\theequation{\arabic{sect}.\arabic{equation}}
                \def\thefigure{\arabic{sect}.\arabic{figure}}}

        \newcommand{\be}{\begin{equation}}
        \newcommand{\ee}{\end{equation}}
        \newcommand{\bea}{\begin{eqnarray}}
        \newcommand{\eea}{\end{eqnarray}}
        \newcommand{\nno}{\nonumber \\}
        \newcommand{\sep}[1]{\!\!\!\! &{#1}& \!\!\!\! }
        \newcommand{\eq}{\sep{=}}

        \newcommand{\ii}{\sqrt{-1}\,}

        \newcommand{\e}[1]{e^{{#1}}}
        
        \newcommand{\hf}{\frac{1}{2}}
	\newcommand{\ve}{{\scriptscriptstyle \vee}}
	\newcommand{\we}{{\scriptscriptstyle \wedge}}
        
	\newcommand{\Sum}[1]{\underset{{}^{#1}}{\sum}}
        \newcommand{\bra}{\langle}
        \newcommand{\ket}{\rangle}
        
        \newcommand{\vol}{{\rm vol}\/}
        \newcommand{\ad}{{\rm ad}\/}
        
        \newcommand{\Hom}{{\rm Hom}\/}

        \newcommand{\loong}{_{\rm long}}
        \newcommand{\short}{_{\rm short}}
        
        \newcommand{\SL}{{\rm SL}}
        \newcommand{\SU}{{\rm SU}}
        \newcommand{\set}[2]{\{{#1}|{#2}\}}
	\newcommand{\sqrtn}{\sqrt{n_\g}}
        
        \newcommand{\LZZ}{{}^L\!\ZZ}
        \newcommand{\LLam}{{}^L\!\Lam}

\begin{document}
\begin{flushright}
{\tt arXiv:yymm.nnnn [hep-th]}
\end{flushright}

\vspace{5pt}

        \begin{center}
{\Large\bf $S$-duality in Vafa-Witten theory for non-simply laced gauge
groups}\\
        \vspace{4ex}
        {\large\rm Siye Wu}

	\vspace{2ex}

{\small {\em Department of Mathematics, University of Hong Kong, Pokfulam,
 Hong Kong, China}\footnote{Current address. E-mail: {\tt swu@maths.hku.hk}}

and

      {\em Department of Mathematics, University of Colorado, Boulder,
        CO 80309-0395, USA}}
        \end{center}

       \vspace{2ex}

        \begin{quote}
{\small 
\noindent {\bf Abstract.}
Vafa-Witten theory is a twisted $N=4$ supersymmetric gauge theory whose
partition functions are the generating functions of the Euler number of
instanton moduli spaces.
In this paper, we recall quantum gauge theory with discrete electric and
magnetic fluxes and review the main results of Vafa-Witten theory when
the gauge group is simply laced.
Based on the transformations of theta functions and their appearance
in the blow-up formulae, we propose explicit transformations of the
partition functions under the Hecke group when the gauge group is
non-simply laced.
We provide various evidences and consistency checks.\\

}
	\end{quote}

\vskip 1cm

\sect{Introduction}

One of the most fascinating conjectures in quantum field theory is the
Montonen-Olive duality [\ref{MO}] between electricity and magnetism
as the gauge group is exchanged with its dual [\ref{GNO}].
It was realised subsequently that this duality is more likely to hold
in supersymmetric gauge theories [\ref{WO}, \ref{Os}].
If such a theory can be twisted so that part of the supersymmetry
remains on a curved four-manifold, the theory becomes topological
in the sense that the partition function and observables are
topological or smooth invariants.
In such a case, duality can both be tested by known mathematical results
and predict new ones.
A celebrated example is the twisted $N=2$ supersymmetric gauge theory
[\ref{W88}], in which the duality of low energy descriptions [\ref{SW}]
yields a relation between Donaldson and Seiberg-Witten invariants 
[\ref{W94}].
The $N=4$ supersymmetric gauge theory is believed to have exact
electric-magnetic duality, or $S$-duality [\ref{Os}].
It has three inequivalent twists [\ref{Ya}, \ref{Ma}].
One of the two twists in [\ref{Ya}] is the Vafa-Witten theory [\ref{VW}],
whose partition functions are the generating functions of the Euler
number of instanton moduli spaces.
These partition functions depend on the discrete fluxes of 't Hooft 
[\ref{tH}] and they transform under the modular group $\SL(2,\ZA)$.
This sharpened $S$-duality conjecture [\ref{VW}] impose stringent 
constraints on the Euler number of moduli spaces and has been a 
fruitful source of development in both topology and $S$-duality.
Recently, there has been much interest in the third twist [\ref{Ma}]
for its relation to the geometric Langlands programme 
[\ref{KW}, \ref{GW}, \ref{FW}, \ref{W07}].

In this paper, We revisit the Vafa-Witten theory with an emphasis on
the roles of Langlands duals.
While much progress has been made when the gauge group is simply laced, both
in physics [\ref{MNVW}, \ref{LL}, \ref{JS1}, \ref{JS2}] and in mathematics
[\ref{Kl}, \ref{Yo1}, \ref{Yo2}, \ref{LQ}, \ref{Go2}], there has been almost
no attempt in studying the Vafa-Witten theory in the non-simply laced case
(see however [\ref{Ka}]).
The latter differs from the simply laced case in several crucial ways.
First, as the gauge group is distinct from its dual, we need to consider
simultaneously two sets of discrete fluxes and hence two sets of partition
functions.
Second, the $\ZA_2$-duality of electricity and magnetism is not part of
the modular group, but the Hecke group [\ref{GGPZ}, \ref{DFHK}, \ref{AKS}],
which has different relations on the generators.
Based on the transformations of theta functions and their appearance
in the blow-up formulae, we propose explicit transformations of the 
partition functions with various discrete fluxes under the Hecke group.
This would be the counterpart, when the gauge group is non-simply laced,
of the sharpened $S$-duality conjecture of Vafa and Witten [\ref{VW}].

The organisation of the paper is as follows.
In Section~2, we review 't Hooft's discrete electric and magnetic fluxes
[\ref{tH}] and their role in canonical and path integral quantisation.
Given an arbitrary simple gauge group, we choose a subset of permitted
discrete fluxes so that under $S$-duality, the discrete electric 
and magnetic fluxes are interchanged.
In Section~3, we consider Vafa-Witten theory for simply laced gauge groups.
The sharpened $S$-duality conjecture [\ref{VW}] specifies how the partition
functions with various discrete fluxes transform under the modular group.
{}From this and the above selection of discrete fluxes for arbitrary
gauge groups, we deduce the usual $\ZA_2$-duality which exchanges the
gauge group and its Langlands dual.
To compare with the non-simply laced case, we summarise the relevant
mathematical results, especially the blow-up formulae 
[\ref{Yo1}, \ref{VW}, \ref{LQ}, \ref{Ka}], in which the universal factors
contain theta functions constructed from the coroot lattice 
[\ref{Ka}] and the Dedekind eta function.
In Section~4, we study Vafa-Witten theory for non-simply laced gauge groups.
There are two sets of partition functions which transform under the Hecke
group. 
To find the explicit representation, we start from the theta functions
whose transformations under the Hecke group are known [\ref{Wu}].
Their appearance in the blow-up formulae is then used to determine how
the generators of the Hecke group act on the partition functions.
Our result contains new phase factors which are different from those
proposed in [\ref{VW}].
We then explain the consequences on the blow-up formulae and compactification
of moduli spaces.
Finally, we check that the action of the generators indeed defines 
a representation of the Hecke group.
In Section~5, we discuss some possible future directions from this work.
We collect in Appendix~A some facts on Lie algebras, Lie groups, their 
Langlands duals, invariant bilinear forms and Coxeter, dual Coxeter
numbers. 
In Appendix~B, we review the geometry of fractional instanton numbers 
[\ref{VW}] in the presence of discrete fluxes for arbitrary gauge groups.
We also mention some congreunce properties of the signature of 
four-manifolds and the dimention of instanton moduli spaces.

\sect{Gauge theory with discrete electric and magnetic fluxes}

We consider gauge theories on a Riemannian four-manifold $X$ with 
a compact, simple gauge group $G$.
The fields in such a theory are the gauge potential $A$, which is 
a connection on a principal $G$-bundle $P$ and the matter fields,
denoted by $\psi$ collectively, which are sections of various 
associated bundles of $P$.
The action is
\bea
S[A,\psi]\eq
\int_X\left[\frac{1}{e^2}(F\;|*F)+\frac{\ii\th}{8\pi^2}(F\,|\,F)+\cdots\right]
                                              \nno
\eq\int_X\frac{\ii}{4\pi}\left[\bar\tau\,(F^+|F^+)+\tau\,(F^-|F^-)\right]
   +\cdots,
\eea
where the terms with $\psi$ and the coupling to $A$ are omitted.
Here, the pairing on the curvature $2$-form is given by the bilinear form
$(\cdot|\cdot)$ on $\g$ explained in Appendix~A and the wedge product on forms.
The gauge coupling constant $e$ and the $\th$ angle combine as a complex
coupling $\tau=\th/2\pi+4\pi\ii/e^2$ in the upper-half plane.

In quantum theory, we integrate over $A$ and $\psi$.
When $G$ is simply connected, the instanton numbers are summed over in the
path integral in order to be compatible with the Hamiltonian formalism.
When $G$ is not simply connected, there are additional characteristic
classes and the instanton numbers are no longer integers [\ref{VW}].
If $G=G_\ad$, we have $w_2(P)\in H^2(X,\ZZ)$, where $\ZZ$ is the centre of
the universal covering group $\tilde G$, and an instanton number $k$
satisfying (\ref{number}).
Fixing $w_2(P)=v$, the partition function is 
\be
Z_{X,v}(\tau)=\!\!\!\!\sum_{k\in\ZA-\hf(v|v)}\!\!\!\!\frac{1}{\vol(\GG_{k,v})}
  \mathop{\int_{k(P)=k}}_{w_2(P)=v}\!\!\!\!DA\,D\psi\;\e{-S[A,\psi]},
\ee
where $\GG_{k,v}$ is the group of gauge transformations on a bundle $P$
with the prescribed topology.
Henceforth, we omit the subscript $X$ in $Z_{X,v}(\tau)$ unless confusion
occurs.
Because of the fractional instanton numbers, we have
\be
Z_v(\tau+1)=\e{-\pi\ii(v|v)}\,Z_v(\tau).
\ee
To relate to canonical quantisation, we consider the case $X=S^1\times Y$,
where $S^1$ is the time direction and $Y$ is a spatial three-manifold.
We write $v=(a,m)$ according to the decomposition [\ref{KW}, \S 7.1]
\be
H^2(X,\ZZ)\cong H^1(Y,\ZZ)\oplus H^2(Y,\ZZ).
\ee
Following [\ref{tH}], $m$ is called a discrete magnetic flux and an element
$e\in H^1(Y,\ZZ)^\we=\Hom(H^1(Y,\ZZ),U(1))$ is called a discrete electric flux.
For each pair $(e,m)$, the partition function
\be
Z_{e,m}(\tau)=\!\!\!\!\sum_{a\in H^1(Y,\ZZ)}\!\!\!\!e(a)\,Z_{v=(a,m)}(\tau)
\ee
corresponds to some Hilbert space $\HH_{e,m}$ in the Hamiltonian formalism.

For $G=\tilde G$ and $G=G_\ad$, the Hilbert spaces are, respectively,
$\HH_{\tilde G}=\oplus_{e\in H^1(Y,\ZZ)^\we}\HH_{e,0}$ and
$\HH_{G_\ad}=\oplus_{m\in H^2(X,\ZZ)}\HH_{0,m}$ [\ref{KW}, \S 7.1].
The corresponding partition functions are 
\bea
Z_{\tilde G}(\tau)\eq\!\!\!\!\sum_{e\in H^1(Y,\ZZ)^\we}\!\!\!\!Z_{e,m=0}(\tau)
=|\ZZ|^{b_1(Y)}\,Z_{v=0}(\tau),                \\
Z_{G_\ad}(\tau)\eq\!\!\!\!\sum_{m\in H^2(X,\ZZ)}\!\!\!\! Z_{e=0,m}(\tau)
=\!\!\!\!\mathop{\sum_{a\in H^1(Y,\ZZ)}}_{m\in H^2(Y,\ZZ)}\!\!\!\!
Z_{v=(a,m)}(\tau).
\eea
For a general gauge group $G$ with the same Lie algebra $\g$, we choose the
partition function as
\be\label{allG}
Z_G(\tau)=\!\!\!\!\mathop{\sum_{e|_{H^1(Y,\pi_1(G))}=1}}_{m\in H^2(Y,\pi_1(G))}
\!\!\!\! Z_{e,m}(\tau).
\ee
In addition to having a Hilbert space
\be
\HH_G=\!\!\!\!
\mathop{\bigoplus_{e|_{H^1(Y,\pi_1(G))}=1}}_{m\in H^2(Y,\pi_1(G))}
\!\!\!\!\HH_{e,m},
\ee
this prescription has the following two (somewhat related) advantages.
First, the partition function can be written as
\be\label{ZG}
Z_G(\tau)=|Z(G)|^{b_1(Y)}\!\!\!\!
\mathop{\sum_{a\in H^1(Y,\pi_1(G))}}_{m\in H^2(Y,\pi_1(G))}
\!\!\!\! Z_{v=(a,m)}(\tau)
=|Z(G)|^{-1+b_1(X)}\!\!\!\!\sum_{v\in H^2(X,\pi_1(G))}\!\!\!\! Z_v(\tau),
\ee
an expression which is manifestally relativistic.
(See [\ref{VW}, \S 3.3] for a derivation of the factor $|Z(G)|^{-1+b_1(X)}$
without the space-time splitting.)
Second, the restriction on $e$ in (\ref{allG}) is equivalent to 
$e\in H^1(Y,Z(G))^\we$.
By Poincar\'e duality, $H^2(Y,\pi_1({}^LG))\cong H^1(Y,Z(G))^\we$,
$H^1(Y,Z({}^LG))^\we\cong H^2(Y,\pi_1(G))$.
So when $G$ is replaced by its Langlands dual ${}^LG$, the spaces of
$e$ and $m$ are exchanged.
This makes $S$-duality possible.

\sect{Vafa-Witten theory for simply laced gauge groups: modular invariance}

In [\ref{VW}], Vafa and Witten studied $S$-duality in twisted $N=4$
supersymmetric gauge theory.
Such a theory is topological and can be defined on any curved four-manifold
$X$ while maintaining part of the supersymmetry.
With certain vanishing theorems [\ref{VW}, \S 2.4], the partition function
captures the Euler number of instanton moduli spaces.
Recall that the topology of a $G_\ad$-bundle $P$ over a four-manifold $X$
is determined by $w_2(P)\in H^2(X,\ZZ)$ and an instanton number $k(P)$
satisfying (\ref{number}).
Fixing $w_2(P)=v$, the partition function is [\ref{VW}, \S 3]
\be\label{Zv}
Z_v(\tau)=q^{-s}\!\!\!\!
\sum_{k\in\ZA-\hf(v|v)}\!\!\!\!\chi(\,\overline{\MM_{k,v}}\,)\,q^k,
\ee
where $q=\e{2\pi\ii\tau}$ and $\MM_{k,v}=\MM_{k,v}(X)$ is the moduli space
of anti-self-dual instantons on $X$ with the prescribed topology.
This is the generating function of the Euler number of certain 
compactification of $\MM_{k,v}$.
The factor $q^{-s}$ comes from the modification of the action to ensure
$S$-duality in curved space.
Consequently, even when $v=0$ and $k\in\ZA$, $Z_v(\tau)$ is not invariant
under $\tau\mapsto\tau+1$.
As the theory is topological, $s$ is a linear combination of the Euler
number $\chi$ and the signature $\sig$ of $X$.

For simplicity, we assume that $H_1(X,\ZZ)$ has no $|\ZZ|$-torsion as in
[\ref{VW}].
Then $H^2(X,\ZZ)\cong\ZZ^{b_2}$, where $b_i=b_i(X)$ is the $i$th Betti
number of $X$.
When $\g$ is simply laced, the sharpened $S$-duality conjecture of 
Vafa and Witten [\ref{VW}, \S 3.2] is that
\be\label{w}
Z_v\left(-\frac{1}{\tau}\right)=
\pm\frac{1}{|\ZZ|^{b_2/2}}\left(\frac{\tau}{\ii}\right)^{w/2}
\!\!\!\!\!\sum_{u\in H^2(X,\ZZ)}\!\!\!\e{2\pi\ii(v|u)}Z_u(\tau)
\ee
for some modular weight $w$.
We can eliminate the factor $(\tau/\ii)^w$ by defining 
\be\label{hat}
\hat Z_v(\tau)=\eta(\tau)^{-w}Z_v(\tau),
\ee
where $\eta(\tau)=q^{1/24}\sum_{n=1}^\infty(1-q^n)$ is the Dedekind 
eta function.
Then the transformations become
\be\label{c}
\begin{array}{l}
\hat Z_v(\tau+1)=\e{-\pi\ii c/12-\pi\ii(v|v)}\,\hat Z_v(\tau), \\
\displaystyle
\hat Z_v\left(-\frac{1}{\tau}\right)=\pm\frac{1}{|\ZZ|^{b_2/2}}
\!\!\!\sum_{u\in H^2(X,\ZZ)}\!\!\!\e{2\pi\ii(v|u)}\,\hat Z_u(\tau),
\end{array}
\ee
where $c=24s+w$ is also a linear combination of $\chi$ and $\sig$.

The original Montonen-Olive duality conjecture [\ref{MO}] is a consequence
of the sharpened $S$-duality (\ref{c}).
The formula of $\hat Z_v(-1/\tau)$ in (\ref{c}) when $v=0$ already shows the
duality between $\tilde G$ and ${}^L\tilde G=G_\ad$ [\ref{VW}, \ref{KW}].
For any gauge group $G$ with the same Lie algebra $\g$, the partition function
is, according to (\ref{ZG}),
\be
\hat Z_G(\tau)=|Z(G)|^{-1+b_1}\!\!\!\!\sum_{v\in H^2(X,\pi_1(G))}\!\!\!\!
\hat Z_v(\tau).
\ee
Thus we have
\bea
\hat Z_G\left(-\frac{1}{\tau}\right)
\eq|Z(G)|^{-1+b_1}\!\!\!\!\!\!\!\!\sum_{v\in H^2(X,\pi_1(G))}\!\!\!\!\!\!
   \pm|\ZZ|^{-b_2/2}\!\!\!\!\sum_{u\in H^2(X,\ZZ)}\!\!\!\!
   \e{2\pi\ii(u|v)}\,\hat Z_v(\tau) \nno
\eq\pm|Z(G)|^{-\chi/2}\,|\pi_1(G)|^{b_2/2}\!\!\!\!\!\!
   \sum_{v\in H^2(X,\pi_1({}^LG))}\!\!\!\!\!\!\hat Z_v(\tau)          \nno
\eq\pm|Z(G)|^{-\chi/2}\,|Z({}^LG)|^{\chi/2}\;\hat Z_{{}^LG}(\tau).  \label{mo}
\eea
That is, the quantum theory with gauge group $G$ and coupling $-1/\tau$
is the same as that with gauge group ${}^LG$ and coupling $\tau$.
This is the Montonen-Olive duality for a general (simply laced) gauge group.

The constants $c$, $w$, $s$ and the sign in (\ref{c}) are fixed by the 
requirement that (\ref{c}) defines a representation of the modular group 
and by explicit calculations of examples of four-manifolds.
Recall that the modular group $\Gam=SL(2,\ZA)$ is generated by
$T={1\quad 1\choose 0\quad 1}\colon\tau\mapsto\tau+1$ and
$S={0\;\; -1\choose 1\quad 0}\colon\tau\mapsto-1/\tau$ satisfying the relations
\be\label{mod}
S^2=(ST)^3\in Z(\Gam),\quad S^4=I.
\ee
The argument in [\ref{VW}, \S 3.3] for SU($N$) shows that for any simply
laced gauge group, the matrix representations of $T$ and $S$ given by
(\ref{c}),
\be
\TT_{uv}=\e{-\pi\ii c/12-\pi\ii(v|v)}\,\del_{uv},\quad
\SS_{uv}=\pm|\ZZ|^{-b_2/2}\,\e{2\pi\ii(u|v)},
\ee
satisfy the relations in (\ref{mod}) if
\be
c=r_\g\,\chi\!\!\mod 4\quad \mbox{ and } \quad\pm=(-1)^{r_\g(\chi+\sig)/4}.
\ee
Here $r_\g(\chi+\sig)/4\in\ZA$ by (\ref{even}) because the Euler number of 
$\overline{\MM_{k,v}}$ vanishes unless its dimension is even.
In fact, it is believed that if the gauge group is simply laced,
then [\ref{VW}, \ref{LL}, \ref{JS2}]
\be\label{consts}
s=(r_\g+1)\chi/24,\quad w=-\chi,\quad c=r_\g\,\chi.
\ee
This agrees with, for SU(2) and more generally for $\SU(N)$, the calculations
of K3 [\ref{Go1}, \ref{VW}, \ref{MNVW}], $\CO P^2$ [\ref{Kl}, \ref{Yo1}],
$\hf$K3 (rational elliptic surfaces) [\ref{MNVW}, \ref{Yo2}] and rational
surfaces [\ref{Go2}].
It also agrees with the physics calculation of K\"ahler surfaces whose
canonical divisor is a disjoint union of smooth curves [\ref{VW}, \ref{LL}].
For other types of simply laced gauge groups, the partition functions have
also been studied for K3 and $T^4/\ZZ_2$ [\ref{JS1}, \ref{JS2}].

The transformations (\ref{w}), (\ref{c}) with (\ref{consts}) is also 
consistent with the blow-up formulae.
Let $X$ be an algebraic surface and $\tilde X$, its blow-up at a point.
Topologically, $\tilde X$ is the connected sum of $X$ and $\overline{\CO P^2}$.
Thus $H^2(\tilde X,\ZA)\cong H^2(X,\ZA)\oplus H^2(\overline{\CO P^2},\ZA)$,
where $H^2(\overline{\CO P^2},\ZA)$ has one generator $e$ with the pairing 
$e^2=-1$.
So the discrete fluxes $\tilde v$ on $\tilde X$ and $v$ on $X$ are related by
$\tilde v=(v,a\otimes e)$, where $a\in\ZZ\cong\Lam^*/\Lam^\ve$.
Blow-up formulae relate the partition functions of the theories on $X$ and 
those on $\tilde X$.
The obvious generalisation of the $\SU(2)$ case [\ref{VW}] to any simply
laced gauge group (see [\ref{LL}] for $\SU(N)$) is
\be\label{fac}
\hat Z_{\tilde X,\tilde v}(\tau)=\hat\th_a(\tau)\,\hat Z_{X,v}(\tau),
\ee
where $\hat\th_a(\tau)=\eta(\tau)^{-r_\g}\th_a(\tau)$ and
\be\label{th}
\th_a(\tau)=\!\!\sum_{x\in\Lam^\ve+a}\!\!\e{\pi\ii(x|x)\tau}.
\ee
Note that $\hat\th_a(\tau)$ ($a\in\ZZ$) are the level $1$ affine characters
[\ref{KP}, \ref{GW}] and transform under the modular group $\Gam$.
The representation of $\Gam$ on $\{\hat Z_{\tilde X,\tilde v}\}$ is
the tensor product of those on $\{\hat Z_{X,v}\}$ and on $\{\hat\th_a\}$.

Mathematically, (\ref{fac}) can be written more explicitly as,
for $\tilde v=(v,a\otimes v)$,
\be\label{blow}
\sum_{k\in\ZA-\hf(\tilde v|\tilde v)}\!\!\!
\chi(\,\overline{\MM_{k,\tilde v}(\tilde X)}\,)\,q^k=
\frac{q^{(r_\g+1)/24}}{\eta(\tau)^{r_\g+1}}\,\th_a(\tau)
\!\!\!\sum_{k\in\ZA-\hf(v|v)}\!\!\!\chi(\,\overline{\MM_{k,v}(X)}\,)\,q^k.
\ee
In fact, the factorisation (\ref{fac}) for $G=\SU(2)$ or $SO(3)$ was motivated
by Yoshioka's work [\ref{Yo1}] (when $X$ is projective and $a=0$) and the
requirement of $S$-duality [\ref{VW}, \S 4.3].
Li and Qin [\ref{LQ}] proved, again when $G=\SU(2)$ or $SO(3)$, that 
(\ref{blow}) holds for any smooth algebraic surface $X$ and $a\in\ZZ$,
if $\overline{\MM_{k,v}}$ is the Gieseker compactification of $\MM_{k,v}$.
The power of the Dedekind eta function in the denominator comes from the
boundary components of the moduli spaces included during compactification.
Ignoring their contributions, Kapranov [\ref{Ka}] showed that 
\be\label{kap}
\sum_{k\in\ZA}\chi(\MM_{k,0}(\tilde X))\,q^k=\th_0(\tau)
\sum_{k\in\ZA}\chi(\MM_{k,0}(X))\,q^k
\ee
for any (possibly non-simply laced) gauge group.
This provides further support of the appearance of $\th_a(\tau)$ in 
the universal factor on the right hand side of (\ref{blow}).

\sect{Vafa-Witten theory for non-simply laced gauge groups: the Hecke group}

For non-simply laced gauge groups, duality exchanges the parameter $\tau$
with $-1/n_\g\tau$ [\ref{GGPZ}, \ref{DFHK}, \ref{AKS}], where $n_\g$ is
the ratio of the squared lengths of long and short roots.
The transformations $T={1\quad 1\choose 0\quad 1}\colon\tau\mapsto\tau+1$
and $S={\quad 0\,\;\;-1/\sqrtn\,\choose\!\!\!\sqrtn\;\;\;\quad 0\;\;\;} 
\colon\tau\mapsto-1/n_\g\tau$ generate the Hecke group 
$G(\sqrtn)\subset\SL(2,\RE)$ and satisfy the relations
\be\label{hecke}
S^2=(ST)^{2n_\g}\in Z(G(\sqrtn)),\quad S^4=1.
\ee
There are two sets of partition functions. 
In addition to $\{Z_u(\tau)\}_{u\in H^2(X,\ZZ)}$ given by (\ref{Zv}), we have
$\{Z_\mu(\tau)\}_{\mu\in H^2(X,\LZZ)}$, where
\be
Z_\mu(\tau)=q^{-\check s}\!\!\!\!\sum_{k\in\ZA-\hf(\mu|\mu)}\!\!\!\!
\chi(\,\overline{\MM_{k,\mu}}\,)\,q^k
\ee
is the partition function of the theory with gauge group $({}^LG)_\ad$ and
discrete flux $\mu\in H^2(X,\LZZ)$.
Here $\LZZ\cong\LLam^*/\LLam^\ve$ is the centre of $\widetilde{{}^LG}$.
The generator $T\colon\tau\mapsto\tau+1$ transforms within each 
of the sets $\{Z_u(\tau)\}$ and $\{Z_\mu(\tau)\}$ while 
$S\colon\tau\mapsto-1/n_\g\tau$ interchanges them.

To find how the Hecke group acts on the two sets of partition functions,
we consider the theta functions on which the action of the Hecke group
is known explicitly [\ref{Wu}].
When $G$ is non-simply laced, besides $\{\th_a(\tau)\}_{a\in\ZZ}$ in
(\ref{th}), there is another set $\{\th_\al(\tau)\}_{\al\in\LZZ}$, where
\be\label{th'}
\th_\al(\tau)=\!\!\!\!\sum_{\xi\in(\LLam)^\ve+\al}\!\!\!\!\e{\pi\ii(\xi|\xi)}.
\ee
These theta functions are different from those in the affine characters
[\ref{KP}, \ref{GW}], which are sums over the lattice generated by the 
long roots and transform under the modular group.
With (\ref{th}) and (\ref{th'}), Poisson summation yields [\ref{Wu}]
\be\label{thS}
\begin{array}{l}
\displaystyle\vth_a\left(-\frac{1}{n_\g\tau}\right)
=\frac{n_\g^{r\loong}/2}{|\ZZ|^{1/2}}\left(\frac{\tau}{\ii}\right)^{r_\g/2}
\sum_{\al\in\LZZ}\e{-2\pi\ii\bra\al,a\ket/\sqrtn}\,\vth_\al(z,\tau),   \\
\displaystyle\vth_\al\left(-\frac{1}{n_\g\tau}\right) 
=\frac{n_\g^{r\short/2}}{|\ZZ|^{1/2}}\left(\frac{\tau}{\ii}\right)^{r_\g/2}
\sum_{u\in\ZZ}\e{-2\pi\ii\bra\al,a\ket/\sqrtn}\,\vth_a(z,\tau).
\end{array}
\ee
Thus we encounter the Hecke group.
Let
\be\label{thth}
\begin{array}{l}
\hat\vth_a(\tau)=\eta(\tau)^{-r\loong}\,
  \eta(n_\g\tau)^{-r\short}\,\vth_a(\tau),\\
\hat\vth_\al(\tau)=\eta(\tau)^{-r\short}\,
  \eta(n_\g\tau)^{-r\loong}\,\vth_\al(\tau),
\end{array}
\ee
where $r\loong$ and $r\short$ are the numbers of long, short simple roots
of $\g$, respectively.
Then [\ref{Wu}]
\be
\begin{array}{l}
\hat\vth_a(\tau+1)=\e{-\pi\ii n_\g r_\g\check h({}^L\g)/12h(\g)+\pi\ii(a|a)}
  \,\hat\vth_a(\tau),\\
\hat\vth_\al(\tau+1)=\e{-\pi\ii n_\g r_\g\check h(\g)/12h(\g)+\pi\ii(\al|\al)}
  \,\hat\vth_\al(\tau),\\
\displaystyle\hat\vth_a\left(-\frac{1}{n_\g\tau}\right)
  =\frac{1}{|\ZZ|^{1/2}}\sum_{\al\in\LZZ}
  \e{-2\pi\ii\bra\al,a\ket/\sqrtn}\,\hat\vth_\al(\tau),\\
\displaystyle\hat\vth_\al\left(-\frac{1}{n_\g\tau}\right)
  =\frac{1}{|\ZZ|^{1/2}}\sum_{a\in\ZZ}
  \e{-2\pi\ii\bra\al,a\ket/\sqrtn}\,\hat\vth_a(\tau).
\end{array}
\ee

The transformations of $Z_v(\tau)$ and $Z_\mu(\tau)$ under $T$ are obvious.
Following (\ref{w}) and (\ref{thS}), we assume that the partition functions
transform under $S$ according to
\be
\begin{array}{l}
\displaystyle
Z_u\left(-\frac{1}{n_\g\tau}\right)=\frac{n_\g^{w\loong/2}}{|\ZZ|^{b_2/2}}
\left(\frac{\tau}{\ii}\right)^{w/2}\!\!\!\!\sum_{\mu\in H^2(X,\LZZ)}\!\!\!
\e{2\pi\ii\bra\mu|u\ket/\sqrtn}\,Z_\mu(\tau),   \\
\displaystyle
Z_\mu\left(-\frac{1}{n_\g\tau}\right)=\frac{n_\g^{w\short/2}}{|\ZZ|^{b_2/2}}
\left(\frac{\tau}{\ii}\right)^{w/2}\!\!\!\!\sum_{u\in H^2(X,\ZZ)}\!\!\!
\e{2\pi\ii\bra\mu|u\ket/\sqrtn}\,Z_u(\tau),
\end{array}
\ee
where $w=w\loong+w\short$.
Since $S$ exchanges two sets of partition functions, there is no need
for the $\pm$ sign that was present in (\ref{w}).
As in (\ref{hat}) and (\ref{thth}), let
\be
\begin{array}{l}
\hat Z_u(\tau)=\eta(\tau)^{-w\loong}\,\eta(n_\g\tau)^{-w\short}\,Z_u(\tau),\\
\hat Z_\mu(\tau)=\eta(\tau)^{-w\short}\,\eta(n_\g\tau)^{-w\loong}\,Z_\mu(\tau).
\end{array}
\ee
Then the transformations under $T$ and $S$ become
\be\label{cc}
\begin{array}{l}
\hat Z_u(\tau+1)=\e{-\pi\ii c/12-\pi\ii(u|u)}\,\hat Z_u(\tau),  \\
\hat Z_\mu(\tau+1)=\e{\pi\ii\check c/12-\pi\ii(\mu|\mu)}\hat Z_\mu(\tau),\\
\displaystyle\hat Z_u\left(-\frac{1}{n_\g\tau}\right)=\frac{1}{|\ZZ|^{b_2/2}}
\!\!\!\sum_{\mu\in H^2(X,\LZZ)}\!\!\!\e{2\pi\ii\bra\mu,u\ket/\sqrtn}
\,\hat Z_\mu(\tau),                                           \\
\displaystyle\hat Z_\mu\left(-\frac{1}{n_\g\tau}\right)=\frac{1}{|\ZZ|^{b_2/2}}
\!\!\!\sum_{u\in H^2(X,\ZZ)}\!\!\!\e{2\pi\ii\bra\mu,u\ket/\sqrtn}
\,\hat Z_v(\tau),
\end{array}                                            
\ee
where $c=24s+w\loong+n_\g w\short$ and
$\check c=24\check s+w\short+n_\g w\loong$ are linear combinations 
of $\chi$ and $\sig$.
We have, just as (\ref{mo}),
\be
\hat Z_G\left(-\frac{1}{n_\g\tau}\right)=|Z(G)|^{-\chi/2}\,
|Z({}^LG)|^{\chi/2}\;\hat Z_{{}^LG}(\tau),
\ee
recovering the original Montonen-Olive duality [\ref{MO}].

In [\ref{VW}, \S 3.3], it was proposed that for non-simply laced groups,
formula (\ref{hat}) holds if $c=c_1(\g)\chi$, where 
$c_1(\g)=\dim\g/(1+\check h(\g))$ is the central charge of the WZW model
at level $1$ [\ref{GW}].
We would like to suggest different values of $c$ and $\check c$ so as
to be compatible with the action of the Hecke group.
We claim that (\ref{cc}) holds with
\be\label{ccn}
c=n_\g r_\g\frac{\check h({}^L\g)}{h(\g)}\chi=(r\loong+n_\g\,r\short)\chi,\quad
\check c=n_\g r_\g\frac{\check h(\g)}{h(\g)}\chi=(r\short+n_\g\,r\loong)\chi.
\ee
As $\chi(\tilde X)=\chi(X)+1$, this is consistent with the factorisation
\be
\hat Z_{\tilde X,\tilde u}(\tau)=\hat\th_a(\tau)\,\hat Z_{X,u}(\tau),\quad
\hat Z_{\tilde X,\tilde\mu}(\tau)=\hat\th_\al(\tau)\,\hat Z_{X,\mu}(\tau),
\ee
where $\tilde u=(u,a\otimes e)$, $\tilde\mu=(\mu,\al\otimes e)$.
The partitions functions of $\tilde X$ transform under the Hecke group in
the tensor product representation of those of $X$ and the theta functions.

With the lack of mathematical examples in the non-simply laced case, it is not
possible to fix the constants $s$, $\check s$, $w\loong$, $w\short$ uniquely.
A possible solution is
\be
w\loong=-\frac{r\loong}{r_\g}\chi,\quad w\short=-\frac{r\short}{r_\g}\chi,
\quad s=\frac{1+r_\g^{-1}}{24}c,\quad\check s=\frac{1+r_\g^{-1}}{24}\check c.
\ee
Then the blow-up formulae are
\be\label{blown}
\begin{array}{l}
\displaystyle\sum_{k\in\ZA-\hf(\tilde u|\tilde u)}\!\!\!
\chi(\,\overline{\MM_{k,\tilde u}(\tilde X)}\,)\,q^k=
\left(\frac{q^{(r\loong+n_\g r\short)/24}}
{\eta(\tau)^{r\loong}\,\eta(n_\g\tau)^{r\short}}\right)^{1+r_\g^{-1}}
\!\!\!\th_a(\tau)\!\!\!\sum_{k\in\ZA-\hf(u|u)}\!\!\!
\chi(\,\overline{\MM_{k,u}(X)}\,)\,q^k,   \\
\displaystyle\sum_{k\in\ZA-\hf(\tilde\mu|\tilde\mu)}\!\!\!
\chi(\,\overline{\MM_{k,\tilde\mu}(\tilde X)}\,)\,q^k=
\left(\frac{q^{(r\short+n_\g r\loong)/24}}
{\eta(\tau)^{r\short}\,\eta(n_\g\tau)^{r\loong}}\right)^{1+r_\g^{-1}}
\!\!\!\th_\al(\tau)\!\!\!\sum_{k\in\ZA-\hf(\mu|\mu)}\!\!\!
\chi(\,\overline{\MM_{k,\mu}(X)}\,)\,q^k.
\end{array}
\ee
While the appearance of the theta functions matches [\ref{Ka}], the 
fractional powers of the eta functions suggest that a more sophisticated
compactification of the moduli spaces is necessary when the group is
non-simply laced.
It is possible\footnote{For example, the first factors on the right 
hand sides of the two equations in (\ref{blown}) can be replaced by 
$q^{(r_\g+1)/24}/\eta(\tau)^{r_\g+1}$ and
$q^{n_\g(r_\g+1)/24}/\eta(n_\g\tau)^{r_\g+1}$, respectively.}
to achieve integer powers at the expense of losing the symmetry between 
$\g$ and ${}^L\g$.
Then the moduli spaces $\MM_{k,u}$ and $\MM_{k,\mu}$ would have to be
compactified differently.

We check that with the choices of $c$ and $\check c$ in (\ref{ccn}),
the matrices 
\be
\begin{array}{l}
\TT_{uv}=\e{-\pi\ii c/12-\pi\ii(v|v)}\,\del_{uv},\quad\quad
\check\TT_{\mu\nu}=\e{-\pi\ii\check c/12-\pi\ii(\mu|\mu)}\,\del_{\mu\nu},\\
\SS_{\mu u}=|\ZZ|^{-b_2/2}\,\e{2\pi\ii\bra\mu,v\ket/\sqrtn}=\check\SS_{u\mu}
\end{array}
\ee
($u,v\in H^2(X,\ZZ)$, $\mu,\nu\in H^2(X,\LZZ)$) from (\ref{cc}) indeed 
satisfy the relations in (\ref{hecke}) for the Hecke group.
For non-simply laced simple Lie algebras, the centre $\ZZ$ is either $\ZA_2$
(for $B_r$ and $C_r$) or trivial (for $F_4$ and $G_2$).
Therefore all $v$ and $\mu$ are two-torsions and $\SS^2$ and $\check\SS^2$ 
are the identity matrix.
We show that $(\check\SS\check\TT\SS\TT)^{n_\g}$ is also the identity matrix. 

First, the contribution of new phase factors in (\ref{cc}) involving 
$c$ and $\check c$ is
\be\label{phase}
(\e{-\pi\ii c/12}\e{-\pi\ii\check c/12})^{n_\g}=
\e{-\pi\ii n_\g(n_\g+1)r_\g\,\chi/12}
\ee 
by using the second identity in (\ref{coxeter}).
If $\g$ is of type $F_4$ or $G_2$, then $\ZZ=\{1\}$ and all the matrices
concerned are scalars.
It is easy to check that (\ref{phase}) is equal to $1$ in both 
cases.\footnote{If $c=c_1(\g)\chi$, then (\ref{phase}) would be 
$\e{4\pi\ii\chi/15}$ for $F_4$ and $\e{3\pi\ii\chi/5}$ for $G_2$.}
If $\g$ is of type $B_r$ or $C_r$, say $C_r$, then the discrete fluxes are
$u=x\otimes\check\lam_s$ and $\mu=y\otimes\check\lam_1$ for some 
$x,y\in H^2(X,\ZA_2)$.
Here $\check\lam_s$ and $\check\lam_1$ are, respectively, the fundamental
coweights corresponding to the spinor representation of $B_r$ and the
defining representation of $C_r$.
(Both representations are miniscule.)
A straightforward calculation yields
\be
(u|u)=r\,x^2/2,\quad(\mu|\mu)=y^2,\quad
\bra\mu,u\ket/\sqrt{2}=x\cdot y\mod 2,
\ee
where the pairing $x\cdot y$ is explained in Appendix B.
Using the Wu formula (\ref{wu}), we have
\bea
(\check\SS\check\TT\SS\TT)_{xy}\eq\e{-\pi\ii r\chi/4}\;2^{-b_2}\!\!\!\!
\sum_{z\in H^2(X,\ZA_2)}\!\!\!\!\e{\pi\ii x\cdot z}\e{-\pi\ii z^2}
\e{\pi\ii z\cdot y}\e{-\pi\ii r\,y^2/2}                     \nno
\eq\e{-\pi\ii(\chi/4+r\,y^2/2)}\;\del_{x+y,w_2}
\eea
and
\be
((\check\SS\check\TT\SS\TT)^2)_{xy}=\e{-\pi\ii r(\chi+w_2^2)/2}\,\del_{xy}.
\ee
The phase factor on the right hand side is $1$ by (\ref{vb}) and (\ref{even}).

\sect{Conclusions}

When the gauge group is non-simply laced, we proposed how the partition
functions of the Vafa-Witten theory transform under the generators of
the Hecke group.
Our approach was to use the known transformation of the theta functions
and their appearance in the blow-up formulae.
As a consistency check, we verified that these transformations indeed
define a representation of the Hecke group on two sets of partition
functions for the gauge group and its Langlands dual.
However, much remains to be done.
One of the important problems is to compute the partition functions,
either mathematically for simple examples of four-manifolds such as K3
and rational surfaces or by mass deformation for K\"ahler surfaces whose
canonical divisor is a disjoint union of smooth curves.
Another is to clarify the meaning of fractional powers of the eta functions
in the blow-up formulae from the appropriate compactification of moduli 
spaces. 
We leave these questions for future exploration.

\bigskip
\newpage
\noindent
{\bf Appendix}

        \setcounter{equation}{0}
        \renewcommand{\theequation}{A.\arabic{equation}}
        \begin{flushleft}
{\bf A. Notations and facts of Lie groups and Lie algebras}
        \end{flushleft}

In this paper, $G$ is a simple, connected, compact Lie group
with Lie algebra $\g$.
Let $T$ be a maximal torus of $G$ with Lie algebra $\gt$.
Then $T=\gt/2\pi\ii\ell$ for some lattice $\ell\subset\ii\gt$.
Let $\Del\subset\ii\gt^*$ be the root system and 
$\Del^\ve=\set{\al^\ve}{\al\in\Del}\subset\ii\gt$, the coroot system.
Denote by $\Lam$ and $\Lam^\ve$ the root and coroot lattices,
respectively.
Then the weight and coweight lattices are $(\Lam^\ve)^*$ and $\Lam^*$.
We have the inclusions [\ref{B}, \S IX.4.9]
\be
\Lam^\ve\subset\ell\subset\Lam^*\subset\ii\gt,\quad
\Lam\subset\ell^*\subset(\Lam^\ve)^*\subset\ii\gt^*.
\ee
Let $\tilde G$ be the universal covering group of $G$.
Its centre is $\ZZ\cong\Lam^*/\Lam^\ve$.
The adjoint group $G_\ad=\tilde G/\ZZ$ has $\pi_1(G_\ad)=\ZZ$ and
a trivial centre.
In general, $\pi_1(G)\cong\ell/\Lam^\ve$ and $Z(G)\cong\Lam^*/\ell$.
The maximal tori of $\tilde G$ and $G_\ad$ are, respectively,
$\tilde T=\gt/2\pi\ii\Lam^\ve$ and $T_\ad=\gt/2\pi\ii\Lam^*$.

We fix an inner product $(\cdot|\cdot)$ on $\ii\g$ such that the long
roots are of square length $2$.
Let $n_\g$ be the ratio of square lengths of long and short roots.
$\g$ is simply laced if all roots are of the same length, in which
case we set $n_\g=1$.
Otherwise, $\g$ is non-simply laced and $n_\g$ is either $2$ or $3$.
The Langlands dual ${}^L\g$ of $\g$ is the Lie algebra whose root system
is isomorphic to $\Del^\ve$.
To keep the same normalisation on the inner product, the root system of
${}^L\g$ should be ${}^L\!\Del=n_\g^{-1/2}\Del^\ve$.
Thus its (co)root and (co)weight lattices are
\be
\LLam=n_\g^{-1/2}\Lam^\ve,\quad
(\LLam)^\ve=n_\g^{1/2}\,\Lam,\quad
((\LLam)^\ve)^*=n_\g^{-1/2}\Lam^*,\quad
(\LLam)^*=n_\g^{1/2}\;(\Lam^\ve)^*.
\ee
The Lie algebra ${}^L\g$ determines a simply connected Lie group
$\widetilde{{}^LG}$ whose centre is 
${}^L\ZZ\cong(\LLam)^*/(\LLam)^\ve\cong(\Lam^\ve)^*/\Lam$ and
is isomorphic to $\ZZ^\we=\Hom(\ZZ,U(1))$, the character group of $\ZZ$.
The Langlands dual ${}^LG$ of the group $G$ is defined by specifying
$\pi_1({}^LG)$ as the subgroup of characters on $\ZZ$ that is trivial
on $\pi_1(G)$.
We have $\pi_1({}^LG)\cong Z(G)^\we$ and $Z({}^LG)=\pi_1(G)^\we$.
In particular, ${}^L\tilde G=({}^LG)_\ad$ and
${}^L(G_\ad)=\widetilde{{}^LG}$.

The centre $\ZZ$ is closely related to the miniscule representations of
${}^L\g$. 
A representation of $\g$ is miniscule if the weights form a single orbit
under the Weyl group action.
If so, the highest weight is called a miniscule weight.
A miniscule weight is fundamental, but not conversely.
The miniscule weights are in one-to-one correspondence with the non-zero
elements of $(\Lam^\ve)^*/\Lam$, by sending the weight to the coset it 
represents [\ref{B}, \S VIII.7.3]. 
Thus there is a bijection between the set of miniscule and zero weights 
and ${}^L\ZZ$ [\ref{Wu}].
A representation of $G$ is miniscule if the induced representation of
$\g$ is so.
There is a bijection between the set of miniscule and trivial representations
of $G$ and $\pi_1({}^LG)\cong Z(G)^\we$ [\ref{Wu}, \S 2].

We mention some results related to the normalised inner product
$(\cdot|\cdot)$.
First, the Killing form $\kap(\cdot,\cdot)$, extended complex linearly 
to $\g^\CO$ and restricted to $\ii\gt$, is positive definite.
We have $\kap(x,y)=2\check h(\g)(x|y)$ for any $x,y\in\ii\gt$, where
$\check h(\g)$ is the dual Coxeter number of $\g$.
We recall that the Coxeter number $h(\g)$ of $\g$ satisfies
$|\Del|=r_\g\,h(\g)$, where $r_\g$ is the rank of $\g$ [\ref{B}, \S V.6.2].
Second, we have for any $\al\in\Del$, $y\in\Lam^*$,
$(\check\al|\check\al)=4/(\al|\al)\in2\ZA$ and 
$(\check\al|y)=2\bra\al,y\ket/(\al|\al)\in\ZA$.
Consequently, for any $x\in\Lam^\ve$, $y\in\Lam^*$, we have
\be\label{pairing}
(x|x)\in2\ZA,\quad(x|y)\in\ZA.
\ee
However, when both $x,y\in\Lam^*$, $(x|y)\in\QQ$ is not always an integer.
We denote by $m_\g$ the smallest positive integer $m$ so that $m(x|x)/2\in\ZA$
for all $x\in\Lam^*$.
Since $\Lam^*/\Lam^\ve$ is of order $|\ZZ|$, $m_\g$ divides $2|\ZZ|$.
We list below $m_\g$ together with other data for each type of $\g$.
Here $(m,n)$ denotes the greatest common divisor of two positive integers
$m$ and $n$.
\begin{center}
\begin{tabular}{|c|ccccccccc|}\hline
$\g$ & $A_r$ & $B_r$ & $C_r$ & $D_r$ & $E_6$ & $E_7$ & $E_8$ & $F_4$ & $G_2$ \\
\hline
$n_\g$ & $1$ & $2$ & $2$ & $1$ & $1$ & $1$ & $1$ & $2$ & $3$ \\ \hline
$m_\g$ & $2(r+1)/(2,r)$ & $1$ & $2/(2,r)$ & $8/(4,r)$ & $3$ & $4$ & $1$ & 
$1$ & $1$ \\  \hline
$h(\g)$ & $r+1$ & $2r$ & $2r$ & $r+1$ & $12$ & $18$ & $30$ & $12$ & $6$ \\
\hline
$\check h(\g)$ & $r+1$ & $2r-1$ & $r+1$ & $2r-2$ & $12$ & $18$ & $30$ & $9$ &
$4$ \\ \hline
\end{tabular}
\end{center}

Finally, we relate the Coxeter and dual Coxeter numbers of $\g$
to those of ${}^L\g$.
If $\g$ is simply laced, then ${}^L\g\cong\g$ and $\check h(\g)=h(\g)$.
In general, we have [\ref{Wu}]
\be\label{coxeter}
h({}^L\g)=h(\g),\quad \check h(\g)+\check h({}^L\g)=(1+n_\g^{-1})\,h(\g).
\ee
The first identity follows easily from $|{}^L\!\Del|=|\Del|$ while the
second, from (see for example [\ref{Wu}, \S 4])
\be\label{dualco}
\check h(\g)=(r\loong+n_\g^{-1}r\short)h(\g)/r_\g
\ee
and the same equality for ${}^L\g$.
Here $r\loong$ and $r\short$ are the numbers of long, short simple roots
of $\g$, respectively.
We give a simple proof of (\ref{dualco}).
Using $\sum_{\gam\in\Del}\kap_{\al\gam}\kap_{\gam\beta}=\kap_{\al\beta}$,
where $\kap_{\al\beta}=\kap(\al,\beta)$ for $\al,\beta\in\Del$, 
the trace of the matrix $(\kap_{\al\beta})_{\al,\beta\in\Del}$ is
$\sum_{\al\in\Del}\kap_{\al\al}=r_\g$ [\ref{Br}].
This implies that
\be 
|\Del\loong|+n_\g^{-1}|\Del\short|=r_\g\,\check h(\g),
\ee
where $\Del\loong$ and $\Del\short$ are the sets of long and short
roots, respectively.
Since [\ref{B}, exer.\ VI.1.20]
\be
|\Del\loong|=r\loong\,h(\g),\quad |\Del\short|=r\short\,h(\g),
\ee
the result follows.

        \setcounter{equation}{0}
        \renewcommand{\theequation}{B.\arabic{equation}}
        \begin{flushleft}
{\bf B. The geometry of instanton numbers and discrete fluxes}
        \end{flushleft}

Topologically, principal $G$-bundles $P$ over a compact, orientable, 
smooth four-manifold $X$ are classified by $p_1(\ad P)\in H^4(X,\ZA)$
and $w_2(P)=w_2(\ad P)\in H^2(X,\pi_1(G))$. 
The former determines the instanton number 
\be
k(P)=-\bra p_1(\ad P),[X]\ket/2\check h(\g)\in\QQ
\ee
and the latter is a discrete flux [\ref{tH}] and is the obstruction to 
lift $P$ to a $\tilde G$-bundle.
If $G$ itself is simply connected, then $k(P)\in\ZA$ and it is the only
characteristic number of $P$.
To get the most general $w_2(P)$, we take $G=G_\ad$.
Since $\pi_1(G_\ad)=\ZZ$, we have a long exact sequence
\be
\cdots\to H^2(X,\Lam^\ve)\to H^2(X,\Lam^*)\to H^2(X,\ZZ)
\to H^3(X,\Lam^\ve)\to\cdots.
\ee
We assume that all elements in $H^2(X,\ZZ)$ can be lifted to
$H^2(X,\Lam^*)$.
If $\tilde w\in H^2(X,\Lam^*)$ is a lift of $w_2(P)\in H^2(X,\ZZ)$, then there
is a $T_\ad$-bundle $Q\to X$ whose first Chern class is $c_1(Q)=\tilde w$.
We denote by $Q^{-1}$ the $T_\ad$-bundle with $c_1(Q^{-1})=-\tilde w$.
The bundles $P$ and $Q^{-1}$ are both quotients of a 
$\tilde G\times_\ZZ\tilde T$-bundle over $X$ constructed as follows.
Let $\{U_\al\}$ be a good open cover of $X$ and let 
$g_{\al\beta}\colon U_\al\cap U_\beta\to G_\ad$ be the transition functions
of $P$. 
If we lift $g_{\al\beta}$ to 
$\tilde g_{\al\beta}\colon U_\al\cap U_\beta\to\tilde G$, then the functions
$h_{\al\beta\gam}=\tilde g_{\al\beta}\tilde g_{\beta\gam}\tilde g_{\gam\al}
\colon U_\al\cap U_\beta\cap U_\gam\to\ZZ$ form a \v Cech cocycle that 
represents $w_2(P)$.
Let $t_{\al\beta}\colon U_\al\cap U_\beta\to T_\ad$ be the transition
functions of $Q$.
The fact that $c_1(Q)=\tilde w$ is a lift of $w_2(P)$ means that $t_{\al\beta}$
can be lifted to $\tilde t_{\al\beta}\colon U_\al\cap U_\beta\to\tilde T$
so that 
$\tilde t_{\al\beta}\tilde t_{\beta\gam}\tilde t_{\gam\al}=h_{\al\beta\gam}$ 
on $U_\al\cap U_\beta\cap U_\gam$.
The $\tilde G\times_\ZZ\tilde T$-bundle is defined by the transition 
functions $(g_{\al\beta},t_{\al\beta}^{-1})$ modulo the diagonal $\ZZ$-action 
on $\tilde G\times\tilde T$.
If $G=SO(n)$, then $\tilde G\times_\ZZ\tilde T={\rm Spin}^\CO(n)$ and the
existence of a lift $\tilde w$ is equivalent to that of a spin$^\CO$ 
structure on $P$.

Consider now the associated bundle $\ad'Q=Q\times_{T_\ad}\g$, where $T_\ad$
acts on $\g$ by the adjoint representation.
The two vector bundles $\ad'Q$ and $\ad P=P\times_{G_\ad}\g$ have the same
fibre $\g$ and the same $w_2(\ad'Q)=w_2(\ad P)$.
Consequently, they are isomorphic outside one point in $X$ and their
instanton numbers differ by that of a bundle on $S^4$, which is an integer
[\ref{VW}].
To calculate the instanton number of $\ad'Q$, we note that
\be
p_1(\ad'Q)={\textstyle \hf}\,{\rm ch}_2(\ad'Q)=
{\textstyle \hf\Sum{\al\in\Del}}\bra\al,\tilde w\ket^2
=\check h(\g)\,(\tilde w|\tilde w).
\ee
Here $(\cdot|\cdot)$ on $H^2(X,\Lam^*)$ is defined by the normalised inner
product $(\cdot|\cdot)$ on $\ii\gt$ and the intersection form on $H^2(X,\ZA)$.
Thus we have $k(\ad'Q)=-\hf(\tilde w|\tilde w)$.
The number $\hf(\tilde w|\tilde w)\!\!\mod 1$ is independent on the lift 
$\tilde w$ of $w_2(P)$.
In fact, if $\tilde w'$ is another lift, then 
$\tilde w'-\tilde w\in H^2(X,\Lam^\ve)$ and
$\hf(\tilde w'|\tilde w')-\hf(\tilde w|\tilde w)=(\tilde w|\tilde w'-\tilde w)
+\hf(\tilde w'-\tilde w|\tilde w'-\tilde w)\in\ZA$ by (\ref{pairing}).
So we have
\be\label{number}
k(P)=-{\textstyle \hf}\,(w_2(P)|w_2(P))\!\!\!\mod 1.
\ee

An important consequence of (\ref{number}) is that for $G_\ad$-bundles,
instanton numbers are not necessarily integers [\ref{VW}].
For $G_\ad=\SU(n)/\ZA_n$ and $w_2(P)=x\otimes\check\lam_1$, where
$x\in H^2(X,\ZA_n)$ and $\check\lam_1$ is the fundamental (and miniscule)
coweight of $\g$ that corresponds to the defining representation, we have 
$\hf(w_2(P)|w_2(P))=x^2(n-1)/2n\!\!\mod 1$,
which is (3.9) (when $n=2$) and (3.13) of [\ref{VW}].
Here $x^2=\bra x\cup x,[X]\ket$ and $x^2(n-1)/2n$ is well-defined modulo $1$.
The generalisation to arbitrary simply laced groups is straightforward 
[\ref{VW}].
With the proper normalisation $(\cdot|\cdot)$ on $\ii\gt$, we have
(\ref{number}) for non-simply laced Lie groups as well.
Since it is possible to glue instantons [\ref{T}] on $S^4$ to $X$ without
affecting $w_2$, one can exhaust all numbers satisfying (\ref{number})
by choosing various $G_\ad$-bundles with a fixed $w_2$.
However, a non-trivial $w_2(P)$ is not always reflected by a fractional
instanton number.
For each type of simple Lie algebra $\g$, the number $m_\g$ in Appendix A
is the smallest positive integer $m$ such that $m\,k(P)$ is always an integer. 
It can be improved to $m_\g/2$ if $X$ is spin and $m_\g$ is even. 

We consider the tangent bundle $TX$.
The second Stiefel-Whitney class $w_2=w_2(X)=w_2(TX)\in H^2(X,\ZA_2)$
always lifts to $H^2(X,\ZA)$.
This means that $X$ is always spin$^\CO$; $X$ is spin if and only if $w_2=0$.
The Wu formula is
\be\label{wu}
x^2=x\cdot w_2\!\!\mod2,
\ee
where for $x,y\in H^2(X,\ZA_2)$, $x\cdot y=\bra x\cup y,[X]\ket$ is defined
modulo $2$.
If $x$ lifts to $H^2(X,\ZA)$, then $x^2=x\cdot x$ is defined modulo $4$.
Moreover, $w_2^2$ is defined modulo $8$.
It is a classical result [\ref{vdB}, \ref{R}] (see [\ref{FK}, \ref{Z}] for
further developments) that the signature $\sig=\sig(X)$ of $X$ satisfies
\be\label{vb}
\sig=w_2^2\!\!\mod 8.
\ee

Finally, let $\MM_{k,v}=\MM_{k,v}(X)$ be the moduli space of anti-self-dual
connections on a $G$-bundle $P\to X$ with instanton number $k$ and discrete
flux $v$.
Its dimension is [\ref{AHS}]
\be\label{dim}
\dim\MM_{k,v}=-2\,\bra p_1(\ad P),[X]\ket-{\textstyle \hf}\dim G\;(\chi+\sig),
\ee
where $\chi=\chi(X)$ is the Euler number of $X$.
Since $\chi+\sig$ is even for any four-manifold and since 
$\dim G=r_\g\!\!\mod2$, the dimension (\ref{dim}) is even if and only if
\be\label{even}
r_\g\,(\chi+\sig)=0\!\!\!\!\mod 4.
\ee

\bigskip

\noindent
{\small {\bf Acknowledgments.}
The author would like to thank M.\ Jinzenji, N.\ C.\ Leung, J.\ Li, 
W.-P.\ Li, V.\ Mathai, S.~Muhki, X.\ Wang and W.\ Zhang for discussions.
This work is supported in part by CERG HKU705407P.}

        \newcommand{\athr}[2]{{#1}.\ {#2}}
        \newcommand{\au}[2]{\athr{{#1}}{{#2}},}
        \newcommand{\an}[2]{\athr{{#1}}{{#2}} and}
        \newcommand{\jr}[6]{{#1}, {\it {#2}} {#3}\ ({#4}) {#5}-{#6}}
        \newcommand{\pr}[3]{{#1}, {{#2}} ({#3})}
        \newcommand{\bk}[4]{{\it {#1}}, {#2}, {#3} ({#4})}
        \newcommand{\cf}[8]{{\it {#1}}, in: {#2}, pp.\ {#3}-{#4}, {#5},
                 {#6}, {#7} ({#8})} 
	\newcommand{\xxx}[1]{{\tt arXiv:{#1}}}

        \vspace{3ex}
        \begin{flushleft}
{\bf References}
        \end{flushleft}
{\small
        \baselineskip=12pt
\begin{enumerate}

\item\label{AKS}
\au{P.\ C}{Argyres} \an{A}{Kapustin} \au{N}{Seiberg}
On $S$-duality for non-simply-laced gauge groups,
{\it JHEP} 06 (2006) 043, \xxx{hep-th/0603048}

\item\label{AHS}
\au{M.\ F}{Atiyah} \an{N}{Hitchin} \au{I.\ M}{Singer}
\jr{Self-duality in four-dimensional Riemannian geometry}
{Proc.\ Roy.\ Soc.\ Lond.}{A362}{1978}{425}{461}

\item\label{vdB}
\au{F}{van der Blij}
\jr{An invariant of quadratic forms mod $8$}
{Indag.\ Math.}{21}{1959}{291}{293}

\item\label{B}
\au{N}{Bourbaki}
\bk{Groupes et alg\`ebres de Lie, Chap.\ IV, V et VI}
{Hermann}{Paris}{1968}; 
\bk{Chap.\ VII et VIII}{Hermann}{Paris}{1975};
\bk{Chap.\ IX}{Masson}{Paris}{1982}

\item\label{Br}
\au{G}{Brown}
A remark on semi-simple Lie algebras,
{\it Proc.\ Amer.\ Math.\ Soc.} 15 (1964) 518

\item\label{DFHK}
\au{N}{Dorey} \au{C}{Fraser} \an{T.\ J}{Hollowood} \au{M.\ A.\ C}{Kneipp}
\jr{$S$-duality in $N=4$ supersymmetric gauge theories with arbitrary gauge 
group}{Phys.\ Lett.}{B383}{1996}{422}{428}, \xxx{hep-th/9605069}

\item\label{FK}
\au{M}{Freedman} \au{R}{Kirby}
\cf{A geometric proof of Rochlin's theorem}
{Algebraic and geometric topology (Stanford Univ., Stanford, Calif., 1976),
Part 2, Proc. Sympos.\ Pure Math., XXXII}{85}{97}{ed.\ R.\ J.\ Milgram}
{Amer.\ Math. Soc.}{Providence, R.I.}{1978}

\item\label{FW}
\an{E}{Frenkel} \au{E}{Witten}
\pr{Geometric endoscopy and mirror symmetry}
{preprint, \xxx{0710.5939 [hep-th]}}{2007}

\item\label{GW}
\an{D}{Gepner} \au{E}{Witten}
\jr{String theory on group manifolds}{Nucl.\ Phys.}{B278}{1986}{493}{549}

\item\label{GGPZ}
\au{L}{Girardello} \au{A}{Giveon} \an{M}{Porrati} \au{A}{Zaffaroni}
\jr{Non-Abelian strong-weak coupling duality in (string-derived) $N=4$
supersymmetric Yang-Mills theories}
{Phys.\ Lett.}{B334}{1994}{331}{338}, \xxx{hep-th/9406128};
\jr{$S$-duality in $N=4$ Yang-Mills theories with general gauge groups}
{Nucl.\ Phys.}{B448}{1995}{127}{165}, \xxx{hep-th/9502057} 

\item\label{GNO}
\au{P}{Goddard} \an{J}{Nuyts} \au{D.\ I}{Olive}
\jr{Gauge theories and magnetic charges}{Nucl.\ Phys.}{B125}{1977}{1}{28}

\item\label{Go1}
\au{L}{G\"ottsche}
\jr{The Betti numbers of the Hilbert scheme of points on a smooth projective
surface}{Math.\ Ann.}{286}{1990}{193}{207}

\item\label{Go2}
\au{L}{G\"ottsche}
\jr{Theta functions and Hodge numbers of moduli spaces of sheaves on rational
surfaces}{Commun.\ Math.\ Phys.}{206}{1999}{105}{136}

\item\label{GuW}
\an{S}{Gukov} \au{E}{Witten}
\pr{Gauge theory, ramification, and the geometric Langlands program}
{preprint, \xxx{hep-th/0612073}}{2006}

\item\label{tH}
\au{G}{'t Hooft}
\jr{On the phase transition towards permanent quark confinement}
{Nucl.\ Phys.}{B138}{1978}{1}{25};
\jr{A property of electric and magnetic flux in nonabelian gauge theories}
{Nucl.\ Phys.}{B153}{1979}{141}{160}

\item\label{JS1}
\an{M}{Jinzenji} \au{T}{Sasaki}
\jr{$N=4$ supersymmetric Yang-Mills theory on orbifold-$T^4/\ZA_2$}
{Mod.\ Phys.\ Lett.}{A16}{2001}{411}{428}, \xxx{hep-th/0012242};
ibid.: higher rank case, {\it JHEP} 12 (2001) 002, \xxx{hep-th/0109159}

\item\label{JS2}
\an{M}{Jinzenji} \au{T}{Sasaki}
An approach to $N=4$ $ADE$ gauge theory on $K3$,
{\it JHEP} 09 (2002) 002, \xxx{hep-th/0203179}

\item\label{KP}
\an{V.\ G}{Ka\v c} \au{D.\ H}{Peterson}
\jr{Infinite-dimensional Lie algebras, theta functions and modular forms}
{Adv.\ Math.}{53}{1984}{125}{264}

\item\label{Ka}
\au{M}{Kapranov}
\pr{The elliptic curve in the $S$-duality theory and Eisenstein series for
Ka\v c-Moody groups}{preprint, \xxx{math.AG/0001005}}{2000}

\item\label{KW}
\an{A}{Kapustin} \au{E}{Witten}
\jr{Electric-magnetic duality and the geometric Langlands program}
{Commun.\ Number Theory Phys.}{1}{2007}{1}{236}, \xxx{hep-th/0604151}

\item\label{Kl}
\au{A.\ A}{Klyachko}
\jr{Moduli of vector bundles and numbers of classes}
{Funct.\ Anal.\ Appl.}{25}{1991}{67}{69}

\item\label{LL}
\an{J.\ M.\ F}{Labastida} \au{C}{Lozano}
\jr{The Vafa-Witten theory for gauge group $\SU(N)$}
{Adv.\ Theor.\ Math.\ Phys.}{3}{1999}{1201}{1225}, \xxx{hep-th/9903172}

\item\label{LQ}
\an{W.-P}{Li} \au{Z}{Qin}
\jr{On blowup formulae for the $S$-duality conjecture of Vafa and Witten} 
{Invent.\ Math.}{136}{1999}{451}{482}, \xxx{math.AG/9805054};
\jr{II: the universal functions}{Math.\ Res.\ Lett.}{5}{1998}{439}{453},
\xxx{math.AG/9805055}

\item\label{Ma}
\au{N}{Marcus}
\jr{The other topological twisting of $N=4$ Yang-Mills}
{Nucl.\ Phys.}{B452}{1995}{331}{345}, \xxx{hep-th/9506002}

\item\label{MNVW}
\au{J.\ A}{Minahan} \au{D}{Nemeschansky} \an{C}{Vafa} \au{N.\ P}{Warner}
\jr{E-strings and $N=4$ topological Yang-Mills theories}
{Nucl.\ Phys.}{B527}{1998}{581}{623}, \xxx{hep-th/9802168} 

\item\label{MO}
\an{C}{Montonen} \au{D.\ I}{Olive}
\jr{Magnetic monopoles as gauge particles?}
{Phys.\ Lett.}{B72}{1977}{117}{120}

\item\label{Os}
\au{H}{Osborn}
\jr{Topological charges for $N=4$ supersymmetric gauge theories}
{Phys.\ Lett.}{B83}{1979}{321}{326}

\item\label{R}
\au{V.\ A}{Rokhlin}
\jr{Proof of Gudkov's hypothesis}{Funct.\ Anal.\ Appl.}{6}{1972}{136}{138}

\item\label{SW}
\an{S}{Seiberg} \au{E}{Witten}
\jr{Electric-magnetic duality, monopole condensation, and confinement in
$N=2$ supersymmetric Yang-Mills theory}{Nucl.\ Phys.}{B426}{1994}{19}{52};
\jr{Erratum}{Nucl.\ Phys.}{B430}{1994}{485}{486}, \xxx{hep-th/9407087}

\item\label{T}
\au{C.\ H}{Taubes}
\jr{Self-dual Yang-Mills connections on non-self-dual $4$-manifolds}
{J.\ Diff.\ Geom.}{17}{1982}{139}{170};
\jr{Self-dual connections on $4$-manifolds with indefinite intersection matrix}
{J.\ Diff.\ Geom.}{19}{1984}{517}{560}

\item\label{VW}
\an{C}{Vafa} \au{E}{Witten}
\jr{A strong coupling test of $S$-duality} 
{Nucl.\ Phys.}{B431}{1994}{3}{77}, \xxx{hep-th/9408074}

\item\label{W88}
\au{E}{Witten}
\jr{Topological quantum field theory}
{Commun.\ Math.\ Phys.}{117}{1988}{353}{386}

\item\label{W94}
\au{E}{Witten}
\jr{Monopoles and four-manifolds}
{Math.\ Res.\ Lett.}{1}{1994}{769}{796}, \xxx{hep-th/9411102}

\item\label{W07}
\au{E}{Witten}
\pr{Gauge theory and wild ramification}
{preprint, \xxx{0710.0631 [hep-th]}}{2007}

\item\label{WO}
\an{E}{Witten} \au{D.\ I}{Olive}
\jr{Supersymmetric algebras that include topological charges}
{Phys.\ Lett.}{B78}{1978}{97}{101}

\item\label{Wu}
\au{S}{Wu}
\cf{Miniscule representations, Gauss sum and modular invariance}
{Proc.\ of the 4th Internat.\ Congr.\ of Chinese Mathematicians, vol.\ I}
{442}{455}{eds.\ L.\ Ji et al.}{Higher Education Press}{Beijing}{2007},
\xxx{0802.2038 [math.RT]}

\item\label{Ya}
\au{J.\ P}{Yamron}
\jr{Topological action from twisted supersymmetric theories}
{Phys.\ Lett.}{B213}{1988}{325}{330}

\item\label{Yo1}
\au{K}{Yoshioka}
\jr{The Betti numbers of the moduli space of stable sheaves if rank $2$ on
$\PP^2$}{J.\ reine angew.\ Math.}{453}{1994}{193}{220}

\item\label{Yo2}
\au{K}{Yoshioka}
\jr{Euler characteristics of $SU(2)$ instanton moduli spaces on rational 
elliptic surfaces}{Commun.\ Math.\ Phys.}{205}{1999}{501}{517},
\xxx{math.AG/9805003}

\item\label{Z}
\au{W}{Zhang}
\jr{Circle bundles, adiabatic limits of $\eta$-invariants and Rokhlin 
congruences}{Ann.\ Inst.\ Fourier (Grenoble)}{44}{1994}{249}{270}

\end{enumerate}}

\end{document}